\documentclass[twocolumn,epjc3]{svjour3}          % twocolumn

\smartqed  % flush right qed marks, e.g. at end of proof

\RequirePackage{graphicx} \journalname{Eur. Phys. J. C}
\def\be{\begin{equation}}
\def\ee{\end{equation}}
\def\bea{\begin{eqnarray}}
\def\eea{\end{eqnarray}}

\begin{document}

\title{Observational constraint on the interacting dark energy models including the Sandage-Loeb test }


\author{Ming-Jian Zhang\thanksref{addr1}
        \and
        Wen-Biao Liu\thanksref{e1,addr1}
}

%\thankstext[$\star$]{t1}{Thanks to the title}
\thankstext{e1}{corresponding author \\  e-mail: wbliu@bnu.edu.cn}

\institute{Department of Physics, Institute of Theoretical Physics,
Beijing Normal University, Beijing, 100875, China\label{addr1} }

\date{Received: date / Accepted: date}

\maketitle

\begin{abstract}

Two types of interacting dark energy models are investigated using
the type Ia supernova (SNIa), observational $H(z)$ data (OHD),
cosmic microwave background (CMB) shift parameter and the secular
Sandage-Loeb (SL) test. We find that the inclusion of SL test can
obviously provide more stringent constraint on the parameters in
both models. For the constant coupling model, the interaction term
including the SL test is estimated at $\delta=-0.01 \pm 0.01
(1\sigma) \pm 0.02 (2\sigma)$, which has been improved to be only a
half of original scale on corresponding errors. Comparing with the
combination of SNIa and OHD, we find that the inclusion of SL test
directly reduces the best-fit of interaction from 0.39 to 0.10,
which indicates that the higher-redshift observation including the
SL test is necessary to track the evolution of interaction. For the
varying coupling model, we reconstruct the interaction $\delta (z)$,
and find that the interaction is also negative similar as the
constant coupling model. However, for high redshift, the interaction
generally vanishes at infinity. The constraint result also shows
that the $\Lambda$CDM model still behaves a good fit to the
observational data, and the coincidence problem is still quite
severe. However, the phantom-like dark energy with $w_X<-1$ is
slightly favored over the $\Lambda$CDM model.

\end{abstract}

\section{Introduction}
\label{introduction}

The accelerating expansion of the universe is an extraordinary
discovery of modern cosmology following Hubble's discovery of the
expansion. A number of independent cosmological probes over the past
decade have supported this phenomenon. Examples include observations
of type Ia supernova (SNIa) \cite{riess1998supernova}, large scale
structure \cite{tegmark2004cosmological}, and cosmic microwave
background (CMB) anisotropy \cite{spergel2003first}. After this
discovery, several theoretical attempts have been made to explain
it. They generally include the dark energy, modified gravity and the
local inhomogeneous model. Among the numerous candidates of dark
energy, the $\Lambda$CDM model with a cosmological constant is
considered to be the simplest and most robust from the view of
observations. However, the theoretical magnitude of this constant
from the particle physical theory is of about 120 orders larger than
the constraint from observations. As a result, the so trivial
cosmological constant falls into the entanglement of two notable
problems. One is the fine-tuning problem which states why the
observed value of cosmological constant energy density
$\rho_{\Lambda}$ is so small
\cite{weinberg1989cosmological,weinberg2000cosmological}. The other
is the coincidence problem \cite{zlatev1999quintessence} which
states why magnitude order of the inappreciable cosmological
constant is same as the present matter density with the expansion of
universe, i.e., $\Omega_{\Lambda0} \sim \Omega_{m0}$. Generally, we
believe that the evolution of cosmic component energy density should
satisfy $\rho_i \propto a^{-3(1+w_i)}$ during the expansion of our
universe, where $w_i$ is its equation of state and $a$ is the cosmic
scale factor. Thus, energy density of the cosmological constant with
$w_{\Lambda}=-1$ should not change, while energy density of matter
would decrease with $a^{-3}$.  From the observations, however, they
are comparable at present epoch. Some approaches have been raised to
reconcile this problem, such as the odd anthropic principle
\cite{vilenkin2003cosmological,garriga1999cosmological,garriga2001solutions}
and the ``tracker field" model \cite{copeland2006dynamics}. In the
latter approach, dark energy is no longer a constant, but some
scalar fields which are usually in forms of the quintessence
\cite{ratra1988cosmological}, phantom \cite{caldwell2002phantom},
k-essence \cite{armendariz2001essentials}, as well as quintom
\cite{feng2005dark}. Nevertheless, they can not get rid of the
suspicion of fine-tuning of model parameters in such models.
Moreover, nature of the dark energy is still mysterious. An
interesting alternative is the interacting model which assumes an
interaction between matter and dark energy. In this initial
phenomenological form \cite{dalal2001testing}, evolution of the dark
energy density $\rho_{X}$ is assumed to follow a ratio relation,
namely, $\rho_{X} \propto \rho_{m}a^{\xi} $ and $\Omega_{X} \propto
\Omega_{m}a^{\xi} $, where the scaling parameter $\xi$ is a constant
to response severity of the coincidence problem. Specially, this
model can recover to the $\Lambda$CDM and self-similar solutions
\cite{behnke2002description,carroll2001can} for the case $\xi=3$ and
$\xi=0$, respectively. Because the interaction term in this form is
redshift-dependent, this model is usually called the varying
coupling model. Different from the varying model, a constant
coupling model with constant interaction term is also provided
\cite{wang2005can,amendola2007consequences} in which the matter
density maybe not follow the common relationship $\rho_{m} \propto
a^{-3} $. Forms in this model are plump, such as the general type
$\rho_X/\rho_m=f(a)$ \cite{wei2011cosmological} where $f(a)$ is a
function of the scale factor $a$, or the specific interaction term
models \cite{costa2010cosmological}. Observationally, a large amount
of observational data, such as the SNIa, CMB, the baryonic acoustic
oscillation (BAO) and the observational $H(z)$ data (OHD), are
widely used to place constraint on these coupling models. For the
constant coupling model, investigations in Refs.
\cite{guo2007probing,amendola2007consequences} deem that a large
coupling can change evolution of the universe during the
matter-dominated epoch. While for the varying coupling model,
investigations \cite{chen2010using,cao2011testing,salvatelli2013new}
found that SNIa and BAO data cannot provide stringent constraint on
the parameter $\xi$ until inclusion of the CMB data. We note that
the above observations apart from the CMB mainly focus on the
redshift $z<2$. Therefore, a probe at higher redshift is necessary
and expected to  better track evolution of the universe.

In 1962, Sandage\cite{sandage1962change}  proposed a promising
survey named redshift drift to directly probe the dynamics of the
cosmic expansion. In 1998, Loeb  \cite{loeb1998direct} found that
this observation could be achieved by collecting the secular
variation of expansion rate during the evolution of universe from
the wavelength shift of quasar (QSO) Ly$\alpha$ absorption lines.
Therefore, this observation is usually named the Sandage-Loeb (SL)
test. According to the schedule, it would monitor the cosmic
expansion history in the region  $z=2 \sim 5$ where other probes are
inaccessible. For a complement, it is useful for us to revisit the
interacting dark energy models using this test. Recently, Liske et
al. \cite{liske2008cosmic,liske2008elt,liske2009espresso} simulated
some SL data using the Monte Carlo method. From previous works, we
find that it generally produces excellent constraint on the
cosmological models, such as the holographic dark energy
\cite{zhang2007exploring}, modified gravity models
\cite{jain2010constraints}, new agegraphic and Ricci dark energy
models \cite{zhang2010sandage}. More recently, Li et al.
\cite{li2013probing} found that the SL test is able to markedly
break degeneracies between model parameters of $f(R)$ modified
gravity, and $f(T)$ gravity theory, when combined with the latest
observations. More importantly, the SL test could identify the dark
energy model with oscillating equation of state and the models
beyond general relativity with varying gravitational coupling, while
the SNIa is out of ability \cite{moraes2011complementarity}. In this
paper, we would extend the analysis on coupling dark energy models
to a deeper redshift interval by virtue of this test. Following
previous works, we shall concentrate on two common interacting
models: (1) a model with constant interaction term $\delta$
\cite{wang2005can,amendola2007consequences} and (2) a varying
coupling model with term $\delta (z)$ initially proposed by Dalal et
al. \cite{dalal2001testing}.

The paper is organized as follows. In Section \ref{interaction
model}, we introduce the basic equations of the phenomenological
interacting models. In Section \ref{observation}, we illustrate the
constraints from the updated observations.  In Section
\ref{constraint result}, we display the constraint result from
observational data. Finally, we summarize our main conclusion and
present discussion in Section \ref{conclusion}.

\section{Phenomenological interacting models}
\label{interaction model}

Interacting cosmological model is an alternative way to solve the
coincidence puzzle. In this paper, we will consider two fossil
models with interaction between dark matter and dark energy, namely
the constant coupling and varying coupling models. Throughout this
paper, we assume a flat FRW universe with $\Omega_{m} + \Omega_X=1$
and a constant equation of state (EoS) $w_X$ of the dark energy. The
Friedmann equation in such assumptions is
    \be  \label{Friedmann}
    3H^2=8\pi G (\rho_m + \rho_X).
    \ee
The conservation equations for these interacting models should read
    \begin{eqnarray}
\label{matter conserve}
& &\dot{\rho}_m+3H \rho_m=+\Gamma \rho_m\,, \\
\label{dark energy conserve} & &\dot{\rho}_X+3H(\rho_X+p_X)=-\Gamma
\rho_m \,,
   \end{eqnarray}
where $H=\dot{a}/a$ is the Hubble parameter, $\Gamma$ is the
interaction term. The dot denotes the derivative with respect to the
cosmic time. Note that the total energy density is conserved,
although the individual energy density does not obey the
conservation law. For simplicity, we commonly define a dimensionless
interaction term
    \be   \label{delta}
    \delta = \Gamma/H.
    \ee
Generally, the positive $\delta$ ($\delta >0$) denotes an energy
transfer from dark energy to dark matter, while the energy would
transfer from matter to dark energy for $\delta <0$.

\subsection{constant coupling model}
\label{constant model}

\begin{figure}
\begin{center}
\includegraphics[width=6.5cm,height=5cm]{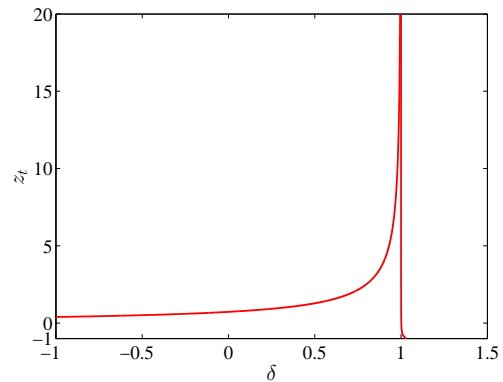}
\end{center}
\caption{ Transition redshift $z_t$ at different interaction term
$\delta$ with fixed $\Omega_{X0}=0.724$ and $w_X=-1.14$ for the
constant coupling model.} \label{transition1}
\end{figure}

For the $\Lambda$CDM model, evolution of the matter energy density
should obey relation $\rho_m \propto a^{-3}$. In order to reconcile
the coincidence problem, the constant coupling dark energy model
states that the evolution of matter density does not satisfy above
relation, but has a small modification to it. Energy density of
matter in this case usually can be written as
\cite{wang2005can,amendola2007consequences,guo2007probing}
    \be \label{matter density}
\rho_m=\rho_{m0}a^{-3+\delta} =\rho_{m0} (1+z)^{3-\delta}\,,
    \ee
where $\rho_{m0}$ is the matter energy density today. The parameter
$\delta$ which should be constrained by the observational data
indicates a deviation of the matter density evolution from regular
relation. Assuming a constant EoS $w_X$ of the dark energy, we
obtain the energy density of dark energy from equation (\ref{dark
energy conserve}) as
      \bea \label{dark energy density}
 \rho_X &=& \rho_{X0} (1+z)^{3(1+w_X)}   \nonumber\\
       &+& \rho_{m0}
\frac{\delta}{\delta+3w_X} \left[
(1+z)^{3(1+w_X)}-(1+z)^{3-\delta}\right],
     \eea
where  $\rho_{X0}$ is the dark energy density today. We note that
the corresponding dark energy density no longer obeys the relation
$\rho_X \propto a^{-3(1+w_X)}$, and presents a decaying component in
the second term of equation (\ref{dark energy density}). The
expansion rate therefore can be obtained following the Friedmann
equation (\ref{Friedmann}) as
    \bea   \label{expansion 1}
    E^2(z) &=& \Omega_{X0}(1+z)^{3(1+w_X)}  \nonumber\\
           &+& \frac{1-\Omega_{X0}}{\delta+3w_X} \left[ \delta
(1+z)^{3(1+w_X)}+3w_X (1+z)^{3-\delta} \right],   \nonumber\\
    \eea
where dark energy density parameter today is $\Omega_{X0} =8 \pi G
\rho_{X0}/(3H_0^2)$. The present matter density parameter is thus
$\Omega_{m0}=1-\Omega_{X0}$. Based on the relationship between
deceleration factor $q(z)$ and expansion rate $E(z)$, the transition
redshift (where $q(z)=0$) from decelerating expansion to
accelerating expansion can be given by
    \be
    z_t = \left[ \frac{3 w_X}{3 w_X +1} \, \frac{(1-\Omega_{X0}) (\delta -1)}{3 w_X \Omega_{X0} + \delta} \,
    \right]^{\frac{1}{3 w_X + \delta}}  - 1.
    \ee
According to the suggestion by WMAP-9 \cite{hinshaw2012nine}, we fix
the present dark energy density parameter $\Omega_{X0}=0.724$,
$w_X=-1.14$ and then plot the transition redshift at different
interaction term $\delta$ in Figure \ref{transition1}. As introduced
in Section \ref{introduction}, accelerating expansion has been
confirmed by many observations. Therefore, the transition redshift
which reflects when acceleration occurs should be positive. In fact,
many literatures found that it may be less than unity. Thus, we
obtain from the Figure \ref{transition1} that interaction term
should be $\delta < 1$. We find that the transition redshift slowly
increases with the increase of $\delta$. Interestingly, a
model-independent transition redshift test can also precisely
determine the $\delta$. For example, as Riess et al.
\cite{riess2004type} evaluated from the SNIa at $z>1$ using the
Hubble Space Telescope, the transition redshift is $z_t=0.46 \pm
0.13$. The corresponding interaction term can be estimated at $-1.21
< \delta < -0.16$.

\subsection{varying coupling model}
\label{vary model}

\begin{figure}
\begin{center}
\includegraphics[width=6.5cm,height=5cm]{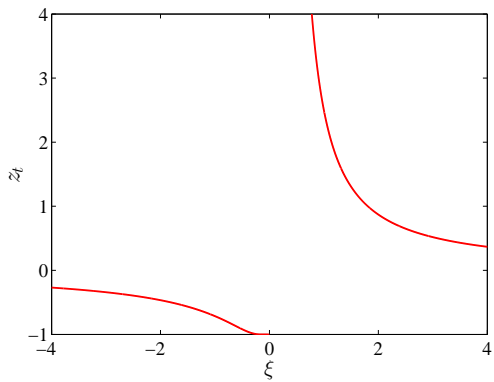}
\end{center}
\caption{Transition redshift $z_t$ at different parameter $\xi$  for
the varying coupling model.} \label{transition2}
\end{figure}

The varying coupling model considered in this section is the
classical scenario proposed by Dalal et al. \cite{dalal2001testing}.
Within the underlying theoretical assumptions, relation between dark
energy and dark matter energy densities is
    \be \label{ratio}
\rho_X \propto \rho_m  a^{\xi}, \qquad  \Omega_X \propto \Omega_m
a^{\xi},
    \ee
where the constant $\xi$ characters severity of the coincidence
problem. Specially, this model can recover to the $\Lambda$CDM and
self-similar solutions \cite{behnke2002description,carroll2001can}
for the case $\xi=3$ and $\xi=0$, respectively. For the FRW universe
with $\Omega_m+\Omega_{X}=1$, the dark energy density parameter
$\Omega_X$ can be solved based on the equation (\ref{ratio}). From
the conservation equations (\ref{matter conserve}) and (\ref{dark
energy conserve}), we can obtain the interaction term
\cite{guo2007probing}
    \be  \label{vary delta}
    \delta(z) = \frac{\delta_0}
 {\Omega_{X0} + (1-\Omega_{X0})(1+z)^\xi}\,,
    \ee
where  $\delta_0 = - (\xi+3w_X)\Omega_{X0}$ is the interaction term
today and $\Omega_{X0}$ is the dark energy density parameter today.
We note that the interaction is absent when $\xi=-3w_X$, which
denotes the standard cosmology. Inversely, the case $\xi \neq -3w_X$
corresponds to a non-standard cosmology. With the interaction term
$\delta$, the dimensionless Hubble parameter can be obtained from
the Friedmann equation (\ref{Friedmann}) \cite{guo2007probing}
    \be   \label{expansion 2}
 E^2(z) = (1+z)^3 \left[1- \Omega_{X0} + \Omega_{X0}(1+z)^{-\xi} \right]^{-3w_X/\xi} \,.
     \ee
The free parameters ($\Omega_{X0}$, $\xi$, $w_X$) eventually can be
determined by the observational data. Following above procedure in
the constant coupling model, we can obtain the corresponding
transition redshift from the deceleration factor $q(z)=0$ as
    \be
    z_t = \left[\frac{1-\Omega_{X0}}{\Omega_{X0} \, (-3w_X -1)}
    \right]^{-1/ \xi} - 1 .
    \ee
Fixing the parameters $\Omega_{X0}$ and $w_X$ suggested by the
WMAP-9, we plot the transition redshift for different $\xi$ in
Figure \ref{transition2}. We find that the positive transition
redshift requires constant $\xi
>0$. With the increase of $\xi$, the transition redshift generally
decreases. In the following section, we will carry out the
observational constraints on these coupling models.

\section{Observational data}
\label{observation}

The observational constraints on the interacting dark energy models
have been performed using the SNIa, BAO, OHD, CMB. The first three
observations mainly focus on the redshift range  $0<z<2$. As a
complement to previous works, we mainly forecast the ability of
future SL test into the deep redshift $2<z<5$. To track the
evolution of interaction over the redshift, we do not use all the
observational data, but only apply the most general SNIa and OHD at
low redshift and the CMB for early epoch as examples, because
previous literatures \cite{guo2007probing,chen2010using} found that
BAO cannot place good constraints on these models.

\subsection{SNIa}
\label{SNIa constraint}

The SNIa data are usually presented as the luminosity distance
modulus. The updated available observation is from the Union2.1
compilation \cite{suzuki2012hubble}, which accommodates 580 data
points. They are discovered by the Hubble Space Telescope Cluster
Supernova Survey over the redshift interval $z<1.415$.
Theoretically, the luminosity distance modulus is usually presented
in the form of the difference between the apparent magnitude $m$ and
the absolute magnitude $M$
    \be   \label{mu}
    \mu_{\scriptsize \textrm{th}}(z)= m - M = 5 \textrm{log}_{10}D_L(z)+\mu_0,
    \ee
where $\mu_0=42.38-5 \textrm{log}_{10} h$, and $h$ is the Hubble
constant $H_0$ in units of 100 km s$^{-1}$Mpc$^{-1}$. The
corresponding luminosity distance function $D_L(z)$  can be
expressed as
\begin{equation}
    \label{DL}
    D_L(z) =  (1 + z) \int^z_0 \frac{\mathrm{d}
    z'}{E(z'; \textbf{p})} ,
\end{equation}
where $\textbf{p}$ stands for the parameters vector of each dark
energy model embedded in expansion rate parameter $E(z';
\textbf{p})$. Commonly, parameters in the expansion rate $E(z';
\textbf{p})$ including the annoying parameter $h$ can be determined
by the general $\chi^2$ statistics. However, an alternative way can
marginalize over the ``nuisance" parameter $\mu_0$
\cite{pietro2003future,nesseris2005comparison,perivolaropoulos2005constraints}.
The remained parameters without $h$ can be estimated  by minimizing
    \be \label{SNchi2}
    \chi^{2}_{\scriptsize \textrm{SN}}(z,\textbf{p})= A-\frac{B^2}{C},
    \ee
where
     \bea
     A(\textbf{p}) &=&  \sum_{i}\frac{[\mu_{\scriptsize \textrm{obs}}(z) - \mu_{\scriptsize \textrm{th}}(z; \mu_0=0,
     \textbf{p})]^2}{\sigma_{i}^{2}(z)},   \nonumber\\
     B(\textbf{p}) &=&  \sum_{i}\frac{\mu_{\scriptsize \textrm{obs}}(z) - \mu_{\scriptsize \textrm{th}}(z; \mu_0=0,
     \textbf{p})}{\sigma_{i}^{2}(z)},   \nonumber\\
     C &=&  \sum_{i} \frac{1}{\sigma_{i}^{2}(z)}.
     \eea
In fact, this program has been widely used in the cosmological
constraints, such as the reconstruction of dark energy
\cite{wei2007reconstruction}, parameter constraint
\cite{wei2010observational},  reconstruction of the energy condition
history \cite{wu2012reconstructing}.

\subsection{OHD}
\label{OHD constraint}

The Hubble parameter $H(z)=\dot{a}/a$ is a key determination in the
research of expansion history of the universe, because it has close
relevance to various observations. In practice, we measure the
Hubble parameter as a function of redshift $z$. Observationally, we
can deduce $H(z)$ from the differential ages of galaxies
\cite{jimenez2008constraining,simon2005constraints,stern2010cosmic},
from the BAO peaks in the galaxy power spectrum
\cite{gaztanaga2009clustering,moresco2012improved} or from the BAO
peak using the Ly$\alpha$ forest of QSOs \cite{2013A&A...552A..96B}.
In addition, we can also theoretically reconstruct $H(z)$ from the
luminosity distances of SNIa using their differential relations
\cite{wang2005uncorrelated,shafieloo2006smoothing,mignone2008model}.
Practically, the available OHD have been applied to constrain the
standard cosmological model
\cite{lin2009observational,stern2010cosmic}, and some other FRW
models
\cite{samushia2008cosmological,zhang2010constraints,zhai2011constraints}.
Interestingly, the potential of future $H(z)$ observations in
parameter constraint has also been explored \cite{ma2011power}. In
this paper, we use the latest available data listed in table 1 of
Ref. \cite{farooq2013hubble}, which accommodates 28 data points.
Parameters can be estimated by minimizing
\begin{equation} \label{OHDchi2}
  \chi^{2}_{\scriptsize \textrm{OHD}}(z, \textbf{p}) = \sum_{i} \frac{[H_0 E(z_i, \textbf{p}) - H^{obs}
  (z_i)]^2}{\sigma_{i}^{2}}.
\end{equation}
In the calculation, we use the Gaussian prior $H_0=70.0 \pm 2.2$ km
s$^{-1}$Mpc$^{-1}$ suggested by the WMAP-9 \cite{hinshaw2012nine}.

\subsection{CMB}
\label{cmb constraint}

The CMB experiment measures the temperature and polarization
anisotropy of the cosmic radiation in early epoch. It generally
plays a major role in establishing and sharpening the cosmological
models. The shift parameter $R$ is a convenient way to quickly
evaluate the likelihood of the cosmological models. For the spatial
flat model, it is expressed as
    \be
    R = \sqrt{\Omega_{m0}}    \int^{z_s}_0 \frac{\mathrm{d}  z'}{E(z'; \textbf{p})} ,
    \ee
where $z_s=1090.97$ is the decoupling redshift
\cite{hu1996small,hinshaw2012nine}. According to the measurement of
WMAP-9, we estimate the parameters by minimizing the corresponding
$\chi^2$ statistics
    \begin{equation}  \label{cmb constraint}
     \chi ^2_R = \left( \frac{R-1.728}{0.016} \right)^2.
     \end{equation}

\subsection{Sandage-Loeb test}
\label{SL constraint}

The Sandage-Loeb (SL) test, namely, redshift drift $\Delta z$ was
first proposed by Sandage \cite{sandage1962change} in 1962. It is a
very potential measurement to directly probe the dynamics of
expansion. In the later decades \cite{loeb1998direct}, many
observational candidates like masers and molecular absorptions were
put forward, but the most promising one appears to be the forest of
the spectra of high-redshift QSOs \cite{pasquini2006codex}. These
spectra are not only immune from the noise of the peculiar motions
relative to the Hubble flow, but also have a large number of lines
in a single spectrum \cite{pasquini2005codex}.  In reality, the
scheduled European Extremely Large Telescope will be equipped with a
high resolution, extremely stable, ultra high precision spectrograph
named the COsmic Dynamics EXperiment (CODEX) that is designed to be
able to measure such signals in the near future.

A signal emitted by a source at time $t_{\mathrm{em}}$ can be
observed at time $t_0$. Because of the expansion of the universe,
the source's redshift should be given through the scale factor
    \be
    z(t_0) = \frac{a(t_0)}{a(t_{\mathrm{em}})} -1.
    \ee
Over the observer's time interval $\Delta t_0$, the source's
redshift  becomes
    \be
    z(t_0+\Delta t_0) = \frac{a(t_0+\Delta t_0)}{a(t_{\mathrm{em}}+\Delta t_{\mathrm{em}})}-1,
    \ee
where $\Delta t_{\mathrm{em}}$ is the time interval-scale for the
source to emit another signal. It should satisfy  $\Delta
t_{\mathrm{em}} = \Delta t_0 / (1+z)$. The observed redshift change
of the source is thus given by
    \be
    \Delta z=\frac{a(t_0+\Delta t_0)}{a(t_{\mathrm{em}}+\Delta t_{\mathrm{em}})}
    - \frac{a(t_0)}{a(t_{\mathrm{em}})}.
    \ee
A further relation can be obtained if we keep the first order
approximation
    \be  \label{change Hubble}
    \Delta z \approx \left[ \frac{\dot{a}(t_0) - \dot{a}(t_{\mathrm{em}})}{a(t_{\mathrm{em}})} \right] \Delta t_0.
    \ee
Clearly, the observable $\Delta z$ is a direct change of the
expansion rate during the evolution of the universe. In terms of the
Hubble parameter $H(z)=\dot{a}(t_{\mathrm{em}})/a(t_{\mathrm{em}})$,
it can be simplified as
   \be  \label{drift def}
   \frac{\Delta z}{\Delta t_0}=(1+z)H_0 - H(z).
   \ee
This is also well known as McVittie Equation
\cite{mcvittie1962appendix}. Taking a standard cosmological model as
an example, we find that the redshift drift at low redshift
generally appears negative with the predominance of matter density
parameter $\Omega_{m0}$. This feature is often regarded as a method
to distinguish dark energy models from void models at $z<2$
(especially at low redshift) \cite{yoo2011redshift}. Unfortunately,
the scheduled CODEX would not be able to measure the drift at such
low $z$, since the target Ly$\alpha$ forest can be measured from the
ground only at $z \geq 1.7$ \cite{liske2008cosmic}. Conveniently, it
is more common to detect the spectroscopic velocity drift
    \be  \label{velocity def}
   \frac{\Delta v}{\Delta t_0}= \frac{c}{1+z} \frac{\Delta z}{\Delta t_0}.
   \ee
It can usually be detected at an order of several cm s$^{-1}$
yr$^{-1}$. Obviously, the velocity variation $\Delta v$ can be
enhanced with the increasing of observational time $\Delta t_0$.

For the capability of CODEX, the accuracy of the spectroscopic
velocity drift measurement was estimated by Pasquini et al.
\cite{pasquini2005codex} using the Monte Carlo simulations
    \begin{equation}\label{velocity error}
\sigma_{\Delta v} = 1.35 \left( \frac{\textrm{S/N}}{2370}
\right)^{-1}\left(  \frac{N_{\scriptsize \textrm{QSO}}}{30}
\right)^{-1/2}\left( \frac{1+z_{\scriptsize \textrm{QSO}}}{5}
\right)^{q} \mathrm{cm/s},
\end{equation}
with $q=-1.7$ for $2<z<4$, or $q=-0.9$ for $z>4$, where S/N is the
signal-to-noise ratio, $N_{\scriptsize \textrm{QSO}}$ and
$z_{\scriptsize \textrm{QSO}}$ are respectively the number and
redshift of the observed QSO. According to currently known QSOs
brighter than 16.5 in  $2<z<5$, we adopt the assumption in Refs.
\cite{pasquini2005codex,whitelock2006scientific} with
$N_{\scriptsize \textrm{QSO}}=30$ and a S/N of 3000. Using the
simulations, the SL test has been widely applied in the model
constraints
\cite{balbi2007time,corasaniti2007exploring,zhang2007exploring,li2013probing}
which is assumed to be uniformly distributed among specific redshift
bins. Following previous works, we would like to examine the ability
of the SL test on the concerned interacting dark energy models. The
mock $\Delta v$ data set for 10 years are assumed to obey uniform
distribution among the redshift bins: $z_{\scriptsize
\textrm{QSO}}=[2 \; 2.5 \; 3 \; 3.5 \; 4 \; 4.5 \; 5]$ in the
fiducial concordance cosmological model. Errors of them can be
calculated from the estimation of equation (\ref{velocity error}).
Parameters of the fiducial $\Lambda$CDM model are given by the
best-fit values of WMAP-9 \cite{hinshaw2012nine}.

\begin{figure}
\begin{center}
\includegraphics[width=6.5cm,height=5cm]{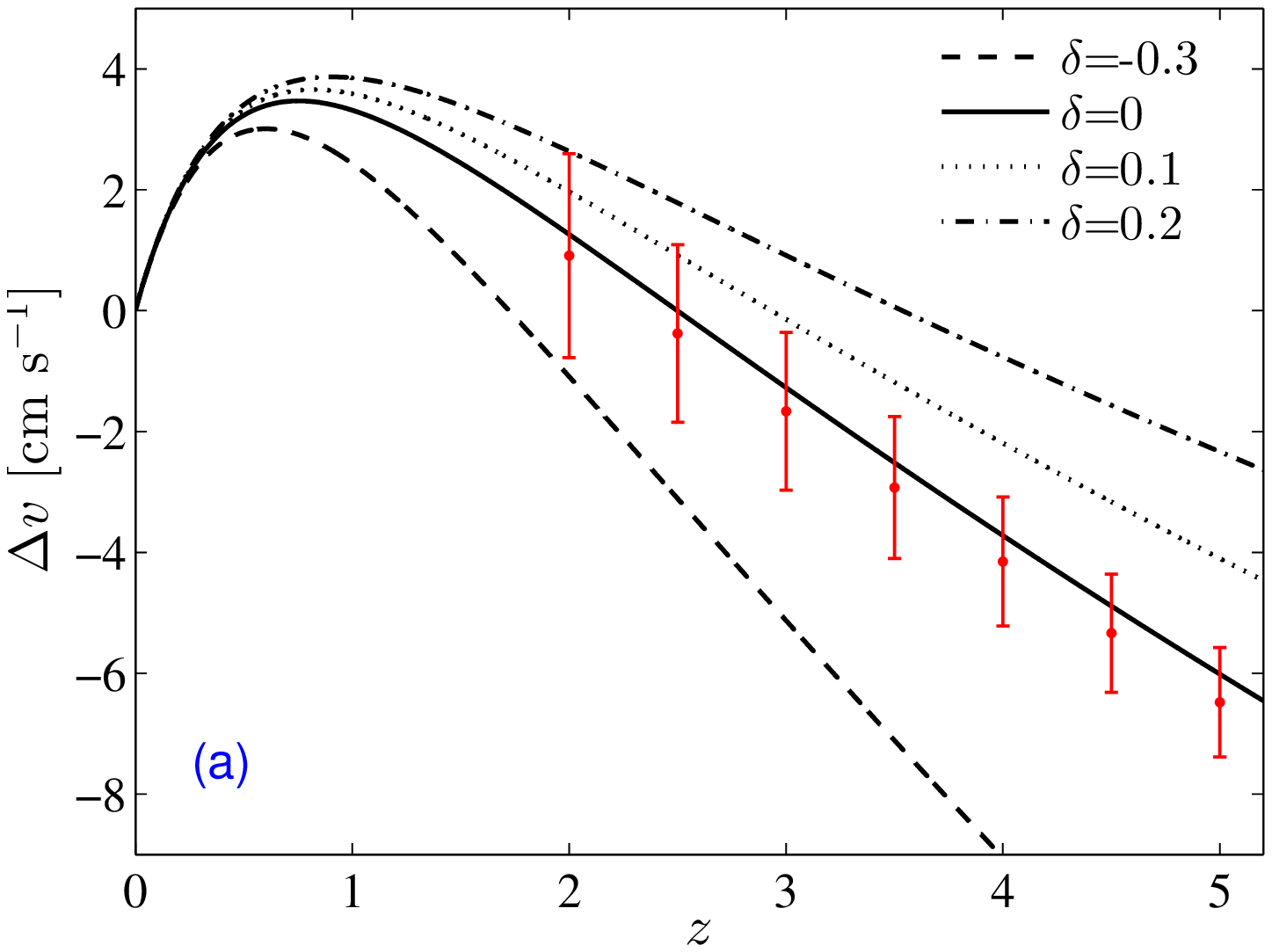}
\includegraphics[width=6.5cm,height=5cm]{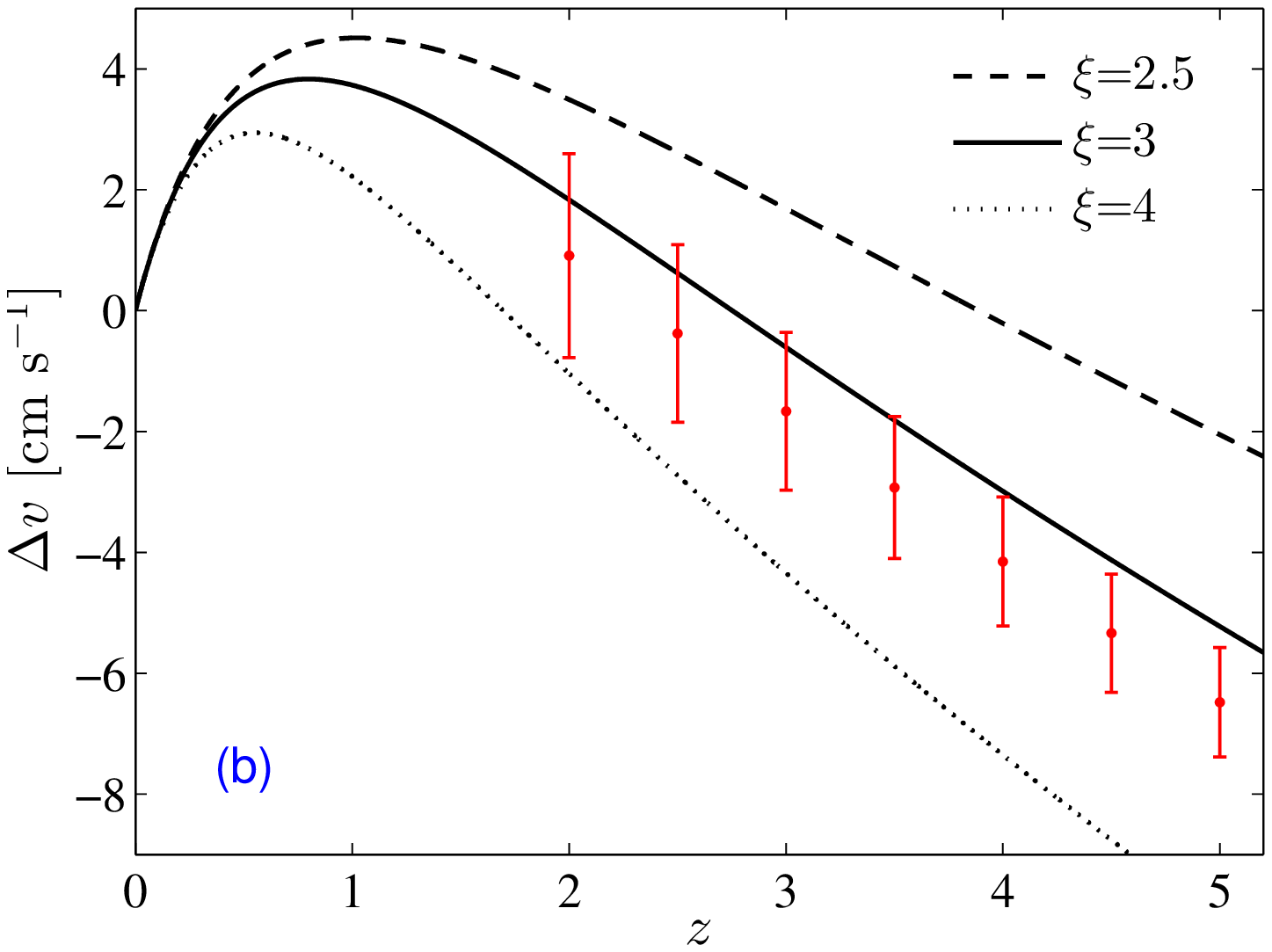}
\end{center}
\caption{Comparison between the simulated
    $\Delta v$ over 10yr observational time and theoretical expectations of the evaluated (a) constant coupling model and (b) varying coupling model for different
    parameters. For the model (a), we change the interaction term $\delta$ and
    fix other parameters under best estimation by Guo et al.
    \cite{guo2007probing}. For the model (b), we change the
    parameter $\xi$ and fix other parameters as best estimation by Cao et al.
    \cite{cao2011testing}. The simulated data points with error bars are estimated by the equation (\ref{velocity error}) in the fiducial model.}
\label{simulation1}
\end{figure}

In Figure \ref{simulation1}, we plot the predicted $\Delta v$ for
different models with different parameters. We find that the
predicted $\Delta v$ curves extend away from each other at high
redshift. In fact, it is useful to precisely determine the
parameters. Comparing with the simulated $\Delta v$, we find that
the parameters are constrained in the narrow regions. For example in
the constant coupling model, if we fix $w_X$ and $\Omega_{X0}$ as
the best estimation by Guo et al. \cite{guo2007probing}, we find
that the interaction term $\delta \sim [-0.3, 0.2]$ seems to be
favored as shown in panel (a). For the varying coupling model, the
parameter $\xi \sim [2.5, 4]$ seems to be favored when we fix other
parameters as the best estimation by Cao et al.
\cite{cao2011testing}. Nevertheless, precise determination of the
parameters should minimize the corresponding $\chi^2$ statistics
    \begin{equation} \label{drift chi2}
  \chi^{2}_{\scriptsize \Delta v}(z, \textbf{p}) = \sum_{i} \frac{[\Delta v^{\scriptsize \textrm{model}}(z_i, \textbf{p}) - \Delta v^{\scriptsize \textrm{data}}
  (z_i)]^2}{\sigma_{\Delta v}^{2}(z_i)},
\end{equation}
where  $\Delta v^{\scriptsize \textrm{model}}(z_i)$ is the
theoretical expectation of the evaluated dark energy models, i.e.,
the constant and varying coupling models. $\Delta v^{\scriptsize
\textrm{data}}(z_i)$ is the mock data produced in the fiducial
$\Lambda$CDM model, and $\sigma_{\Delta v}^{2}(z_i)$ is the
corresponding error estimated by equation (\ref{velocity error}).

In general, we often perform joint analysis by combining several
types of observational data in order to better constrain the
cosmological models. In this paper, we will respectively perform the
likelihood fit by adding different types of observational data in
order to test the evolution of interaction term.

\section{Constraint on the coupling models}
\label{constraint result}

\begin{figure}
\begin{center}
\includegraphics[width=6.5cm,height=5cm]{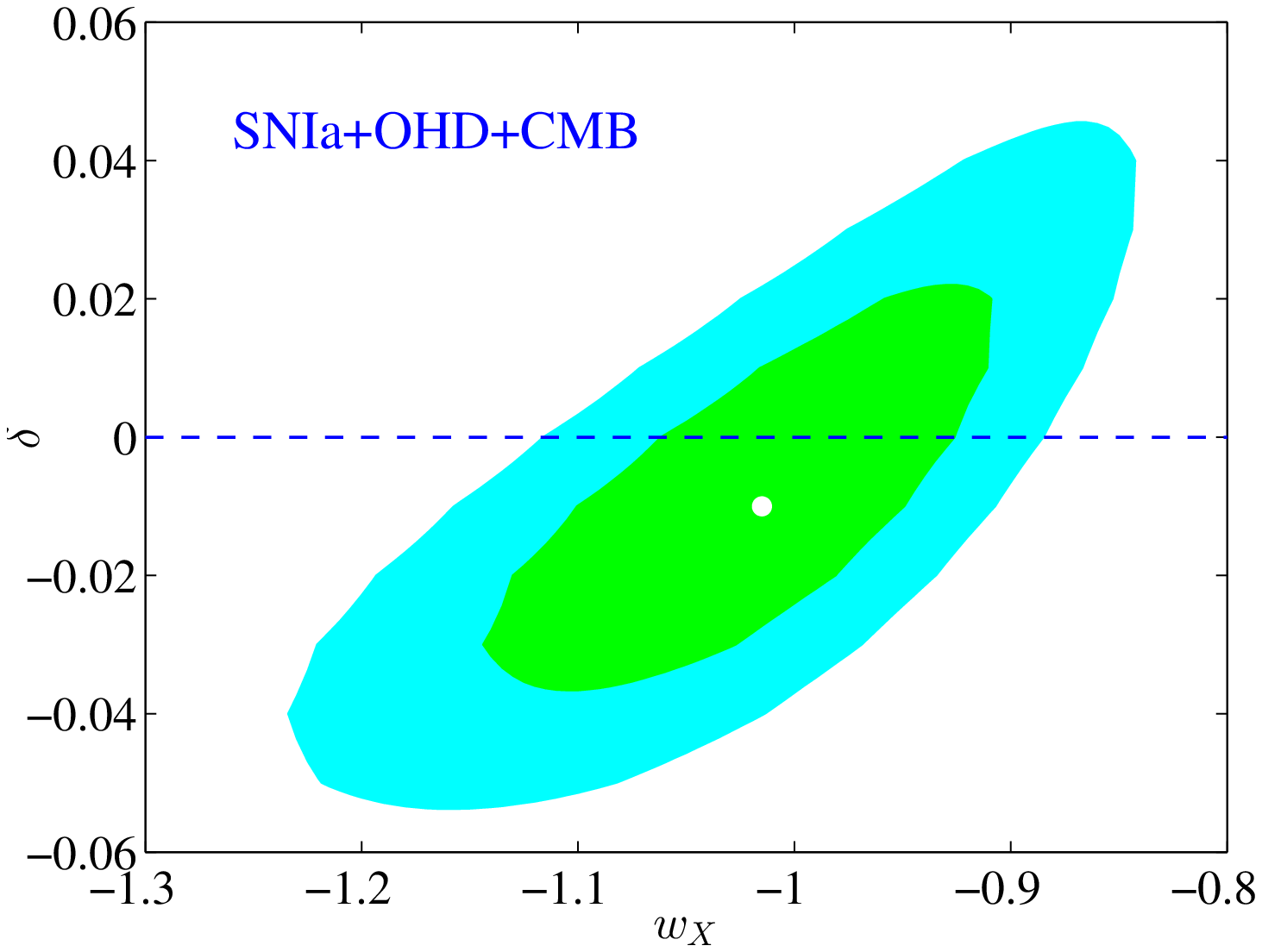}
\includegraphics[width=6.5cm,height=5cm]{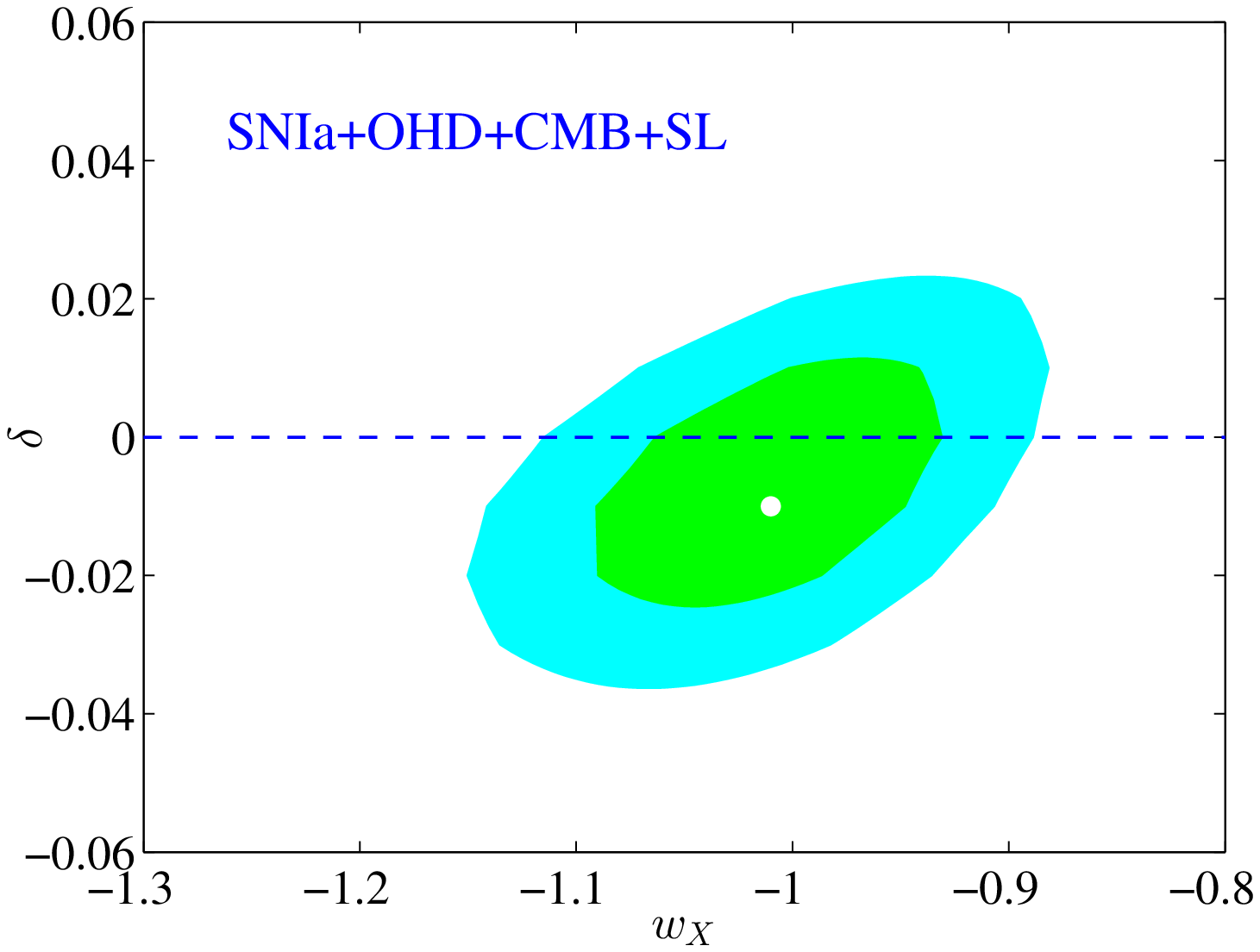}
\includegraphics[width=6.5cm,height=5cm]{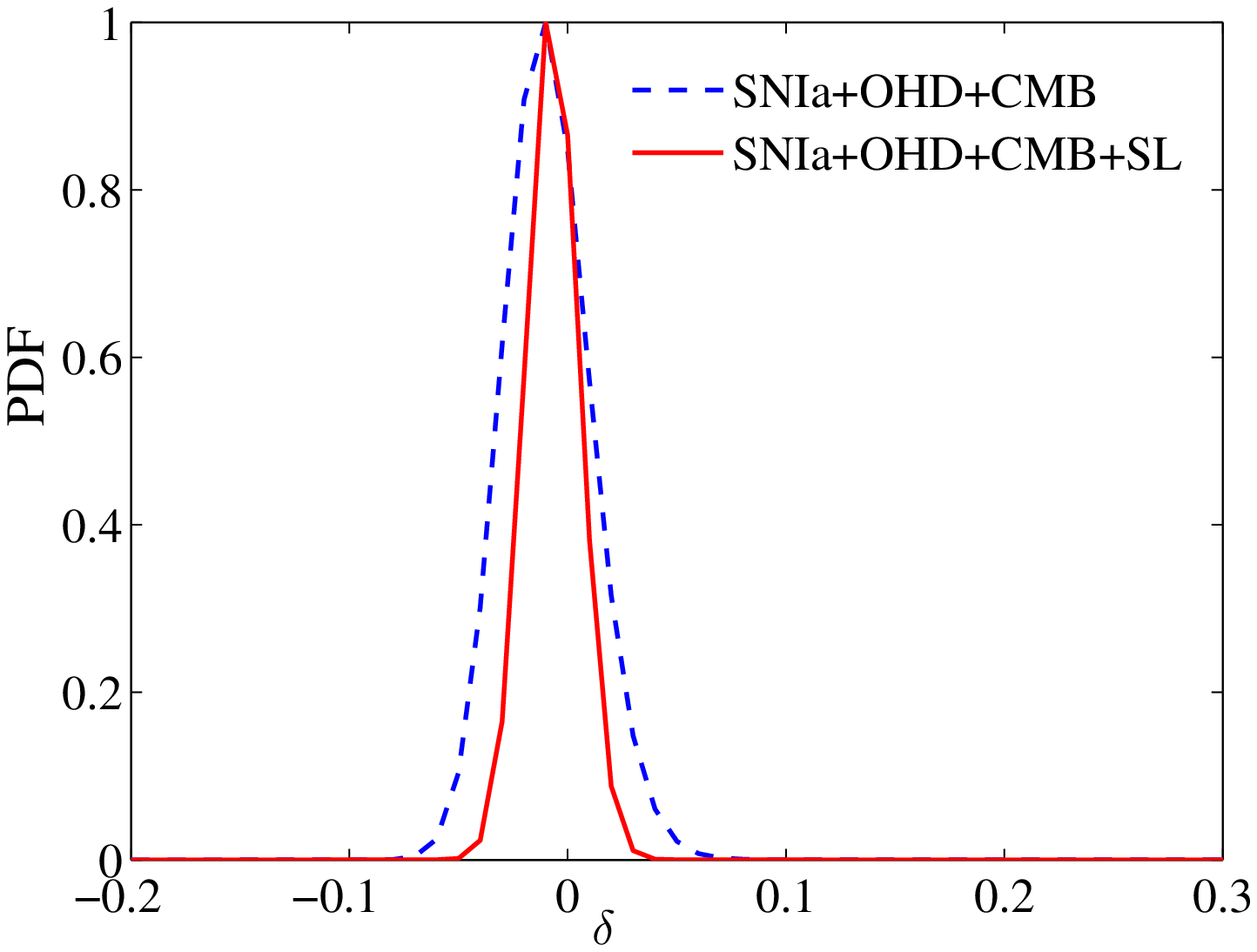}
\end{center}
\caption{Contours correspond to 68.3\%, 95.4\% confidence levels and
the marginalized probability distribution of $\delta$ with different
data sets for the constant coupling model.} \label{constant
contour1}
\end{figure}

\begin{figure}
\begin{center}
\includegraphics[width=6.5cm,height=5cm]{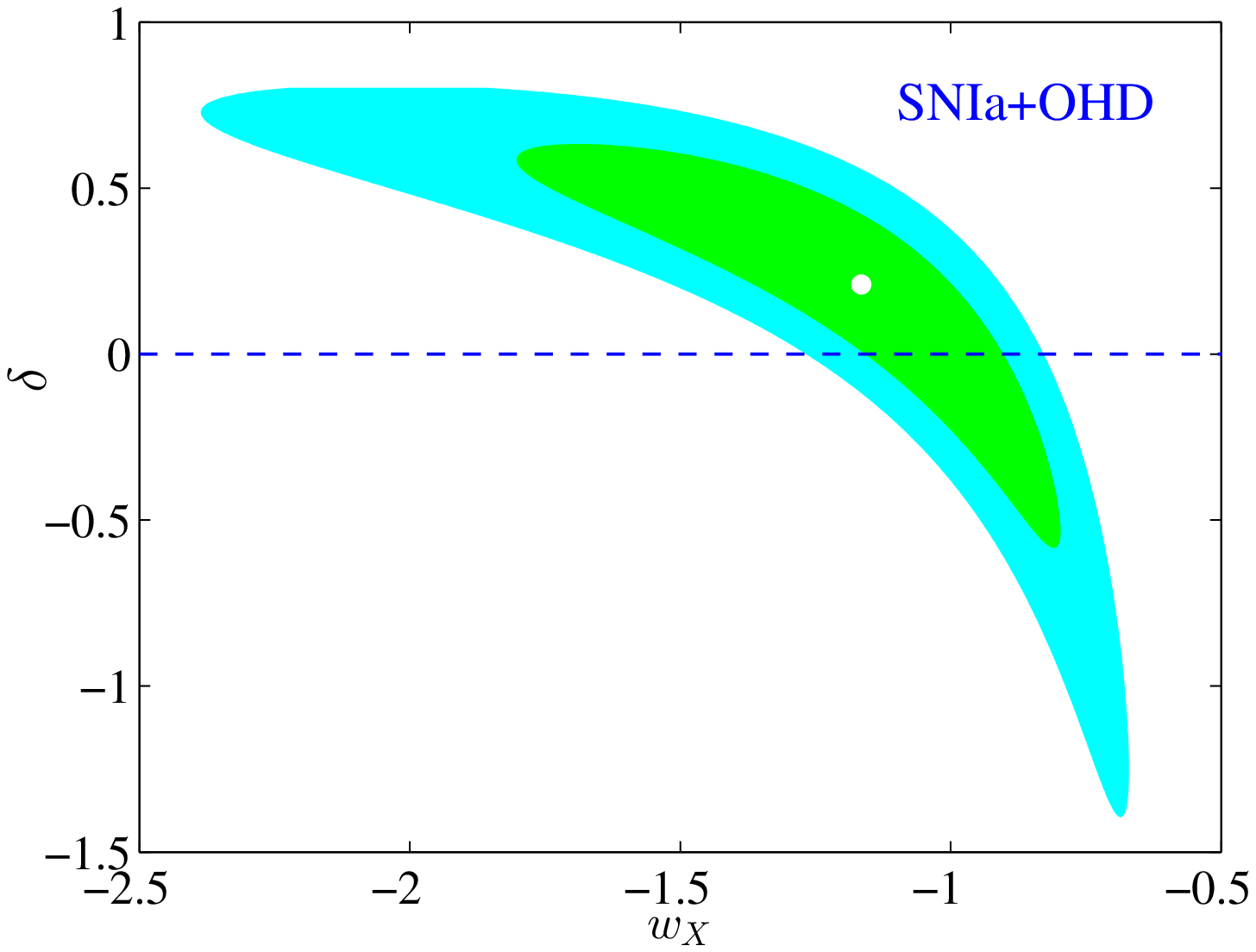}
\includegraphics[width=6.5cm,height=5cm]{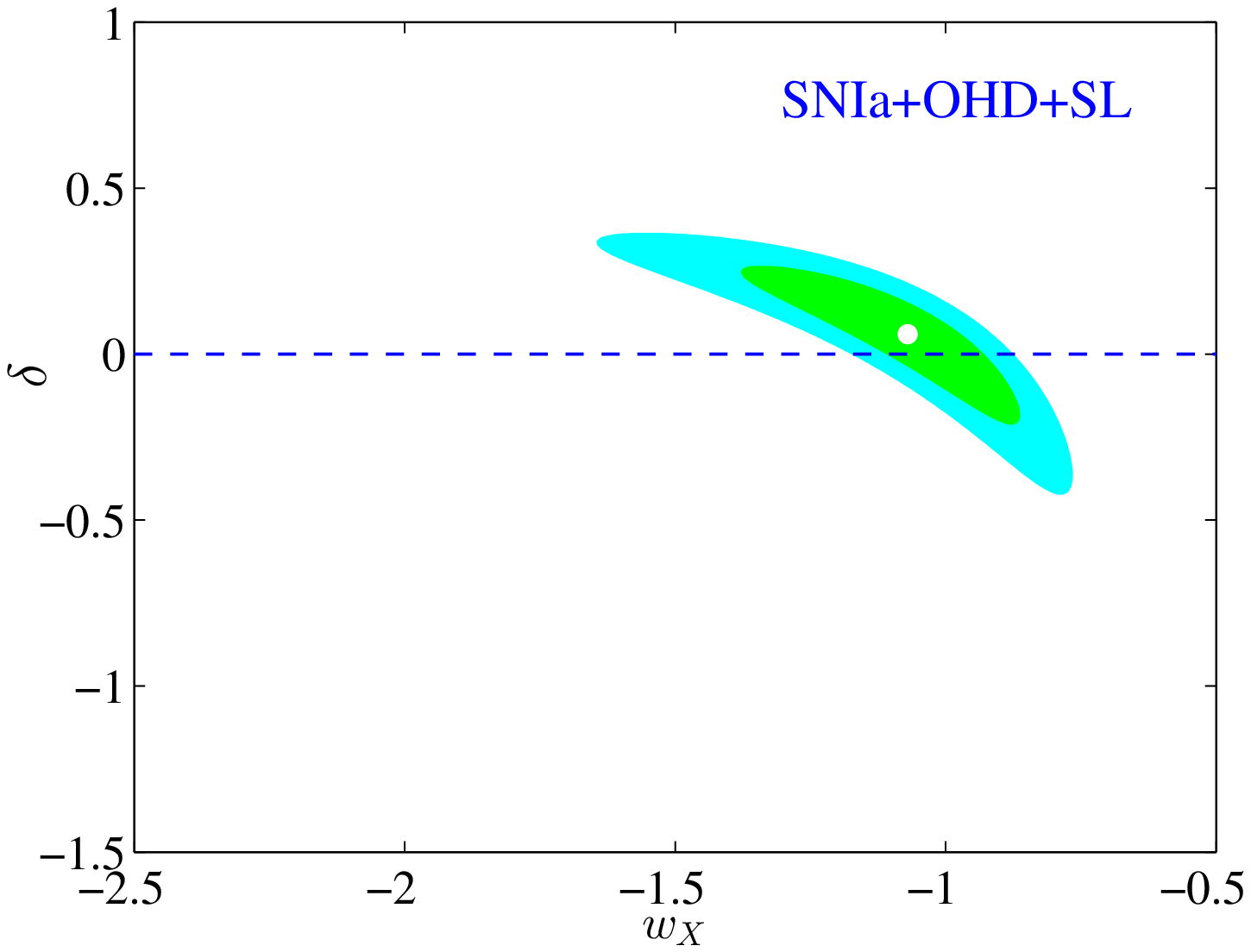}
\includegraphics[width=6.5cm,height=5cm]{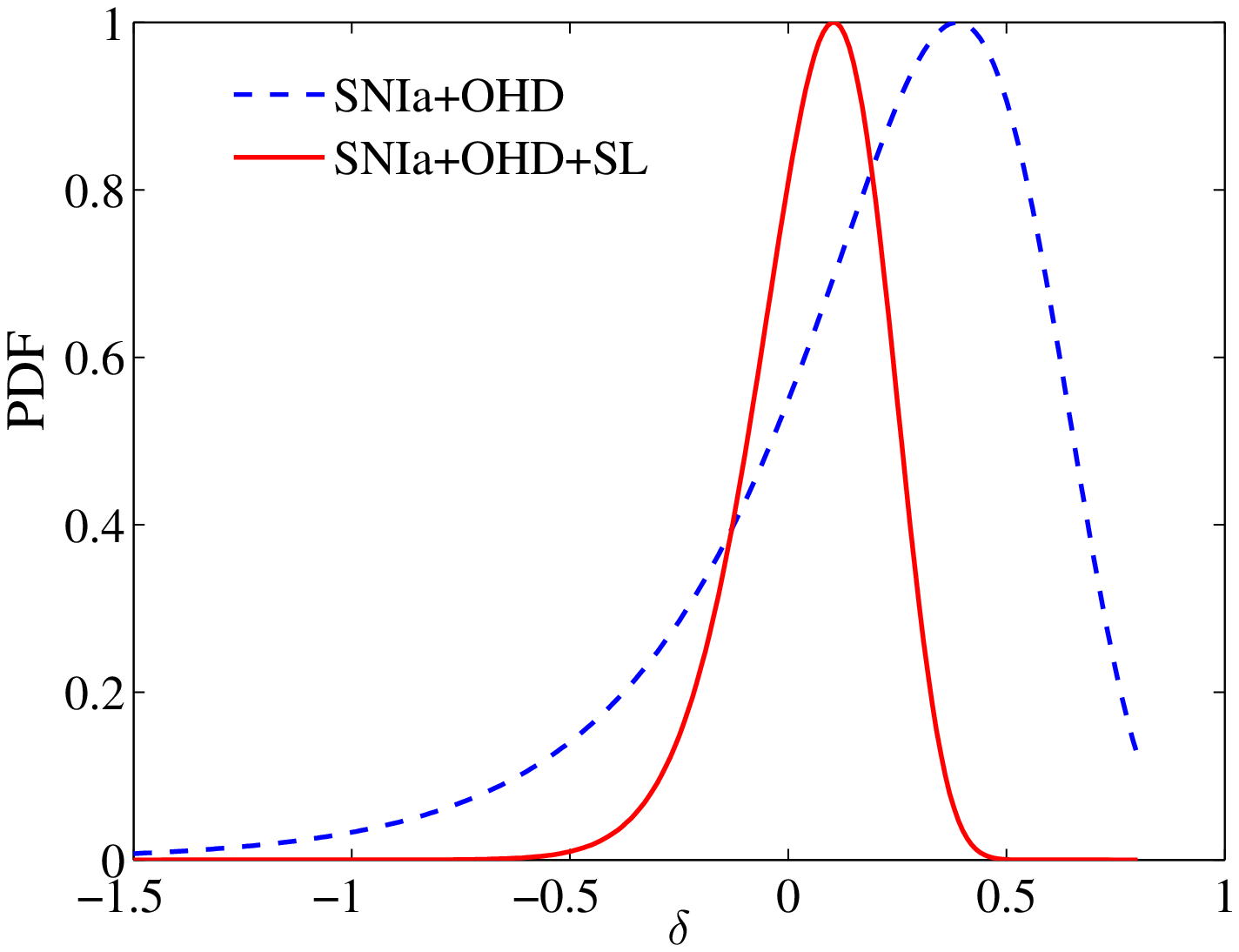}
\end{center}
\caption{Same as Figure \ref{constant contour1} but for different
data sets.} \label{constant contour2}
\end{figure}

By performing the $\chi^2$-test using different data or data sets,
we are able to report the constraint on parameters, and reconstruct
the evolution of interaction term.

For the \emph{constant coupling model}, we implement the likelihood
analysis using different data sets and display the corresponding
contour constraints of parameters ($w_X, \, \delta$) in Figures
\ref{constant contour1} and \ref{constant contour2}, after
marginalizing over the current dark energy density parameter
$\Omega_{X0}$. For the observational data combination SNIa+OHD+CMB,
they give a very severe constraint on the interaction term
$\delta=-0.01 \pm 0.02$($1\sigma$)$\pm 0.04$($2\sigma$), which
presents a weak but negative interaction between dark energy and
matter. The rest parameters EoS  and dark energy density are also
constrained to a good level, $w_X=-1.015^{+0.14}_{-0.17}$($2\sigma$)
and $\Omega_{X0}=0.72^{+0.04}_{-0.04}$($2\sigma$). In order to
forecast the ability of SL test, we add the simulated SL test data
with current data, and show the results in the middle panel of
Figure \ref{constant contour1}. We find that the SL test effectively
reduces the contour constraint region. Especially, the key parameter
interaction term is constrained more stringent, which can be seen
from the marginalized probability distribution (PDF) of $\delta$ in
the right panel. The corresponding constraints are $\delta=-0.01 \pm
0.01$($1\sigma$)$\pm 0.02$($2\sigma$),
$w_X=-1.01^{+0.10}_{-0.11}$($2\sigma$) and
$\Omega_{X0}=0.72^{+0.02}_{-0.01}$($2\sigma$), respectively.
Previous works found that the observational data apart from the CMB
can not constrain the interacting models well. We perform the same
likelihood test from the joint analysis of SNIa and OHD and obtain
$\delta=0.39^{+0.40}_{-0.90}$($2\sigma$), which is much more rough
compared with the inclusion of CMB. As stated by Guo et
al.\cite{guo2007probing}, this is because a large coupling can
change the cosmological evolution during the matter-dominated epoch.
In order to further track the evolution of interaction with
expansion of the universe, we extend our analysis to the higher
redshift using SL test in Figure \ref{constant contour2}. We find
that the inclusion of SL test much improves the constraint,
$\delta=0.10^{+0.25}_{-0.36}$($2\sigma$). The contour constraint
region at $2\sigma$ level with the SL test is even smaller than the
constraint without SL at $1\sigma$ level. The marginalized PDF of
interaction $\delta$ is not only narrowed with high significance,
but also moves towards zero.

\begin{figure}
\begin{center}
\includegraphics[width=6.5cm,height=5cm]{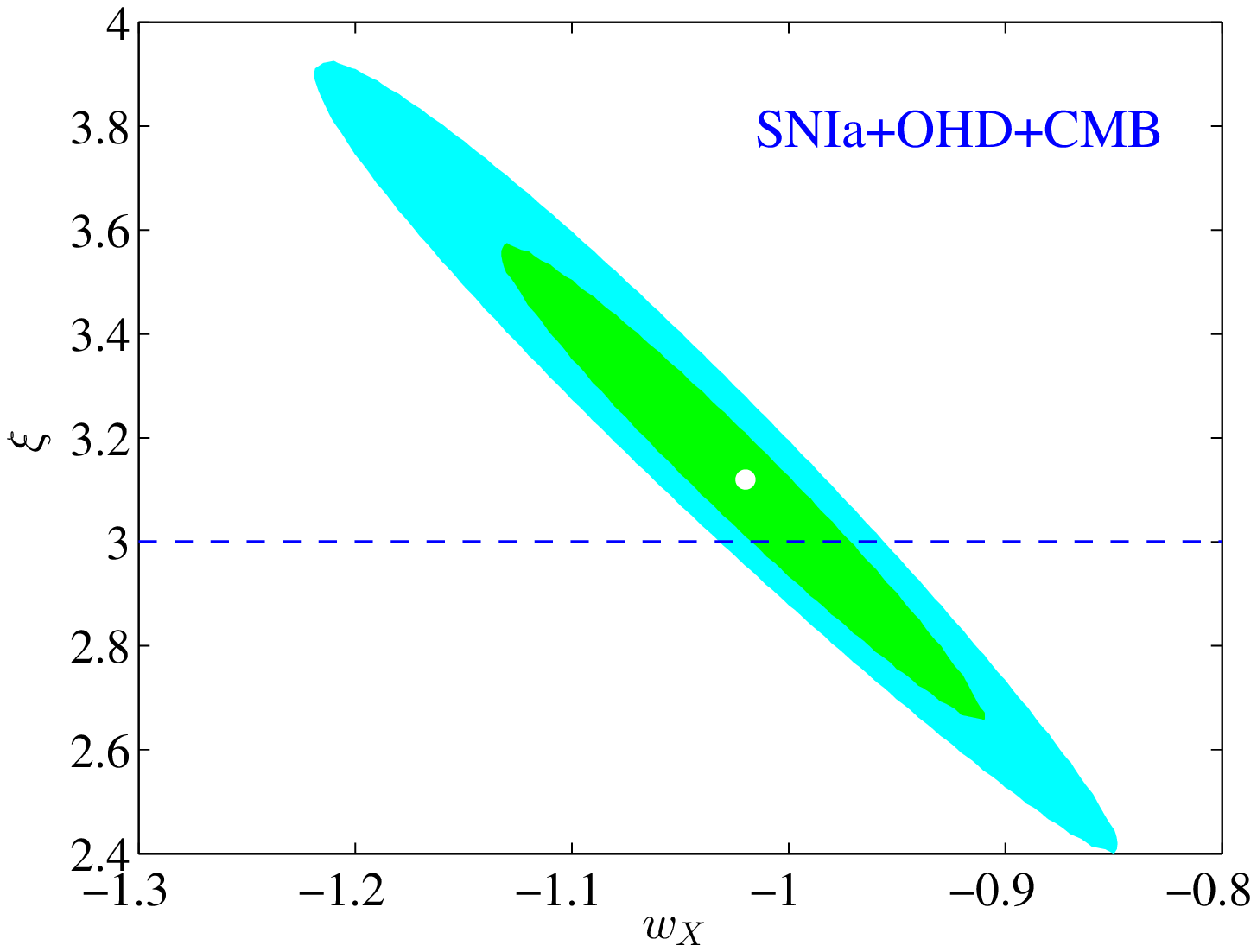}
\includegraphics[width=6.5cm,height=5cm]{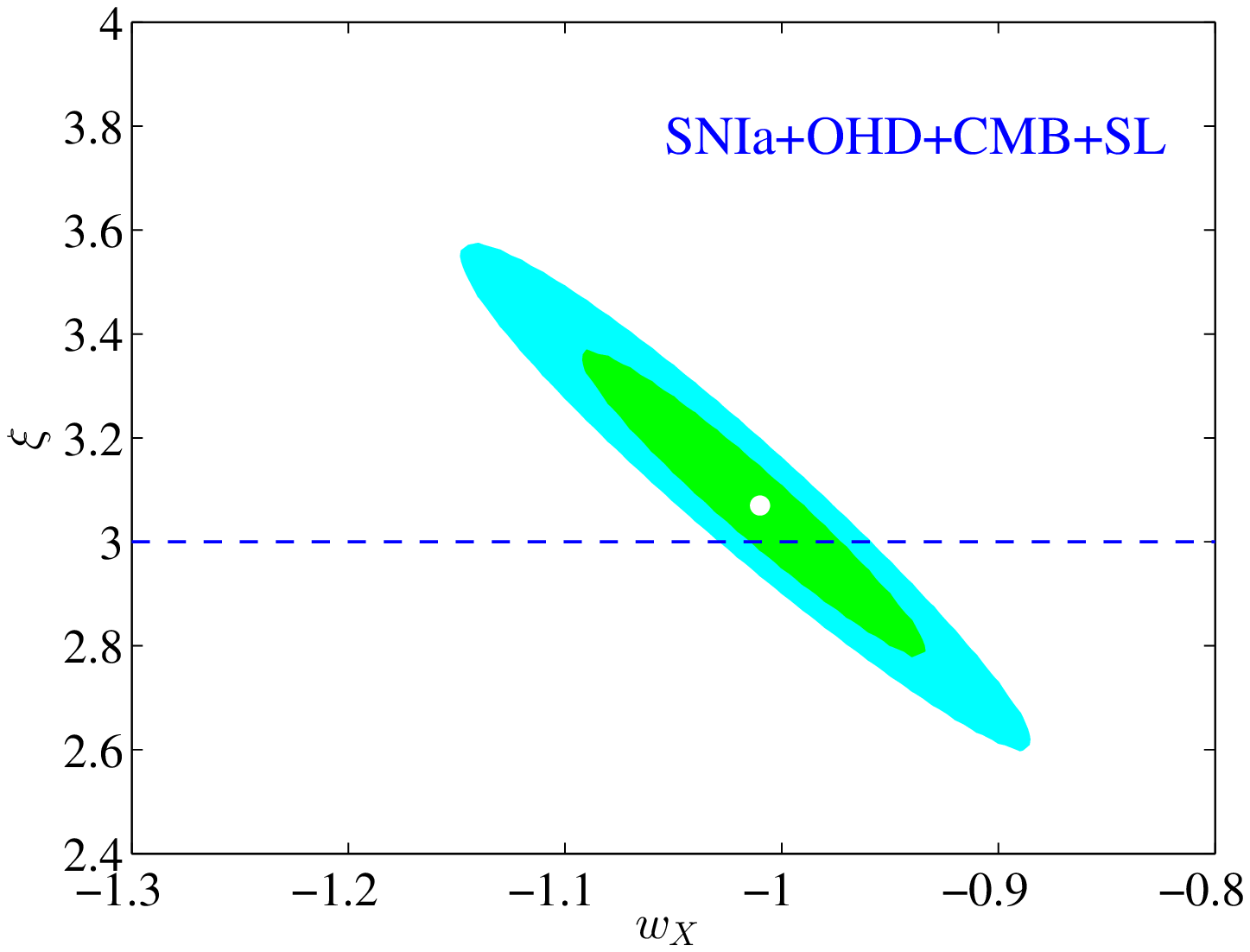}
\end{center}
\caption{ Contours correspond to 68.3\%, 95.4\% confidence levels
for the varying coupling model with different data sets.}
\label{vary contour1}
\end{figure}

\begin{figure}
\begin{center}
\includegraphics[width=6.5cm,height=5cm]{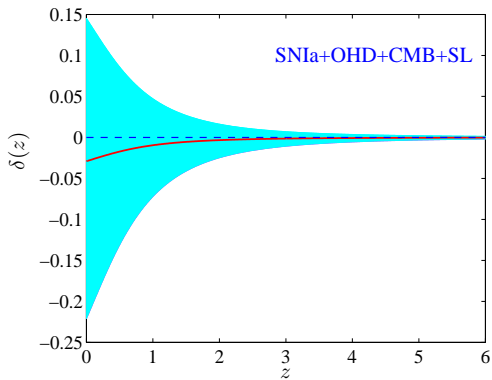}
\end{center}
\caption{ Reconstruction of the interaction term $\delta (z)$ using
total samples of data for the varying coupling model.
    The shaded region corresponds to the errors of $\delta (z)$ at $1\sigma$ C.L.. The red solid curve is the best-fit estimation of $\delta (z)$.}
\label{interaction}
\end{figure}

For the \emph{varying coupling model},  we perform the same
likelihood test using the current observational data with or without
SL test, respectively. For the combination of all considered current
observational data, we find that they can provide fair constraints
on the parameters. For example, the EoS and dark energy density
parameters are $w_X=-1.02^{+0.14}_{-0.16}$ ($2\sigma$),
$\Omega_{X0}=0.72^{+0.04}_{-0.04}$ ($2\sigma$), respectively. The
parameter $\xi=3.12^{+0.31}_{-0.29}$ ($1\sigma$)$^{+0.66}_{-0.57}$
($2\sigma$) represents that the coincidence problem is still severe.
From the equation (\ref{vary delta}), we find that the sign of
interaction term completely depends on the current value
$\delta_0=-(\xi+3 w_X)\Omega_{X0}$. Using the constraint results, we
reconstruct the current interaction $\delta_0=-0.04^{+0.26}_{-0.28}$
($1\sigma$), which presents interaction today is weak but a negative
best-fit value from the available observations. This is consistent
with previous results \cite{guo2007probing}. For forecasting the
power of the SL test, we also include the mock SL data in bottom
panel of Figure \ref{vary contour1}. The allowed region of
parameters is obviously reduced, which is same as the constant
coupling model. The parameters in this case are found to be
$w_X=-1.01^{+0.10}_{-0.11}$ ($2\sigma$),
$\Omega_{X0}=0.725^{+0.02}_{-0.02}$ ($2\sigma$) and
$\xi=3.07^{+0.20}_{-0.19}$ ($1\sigma$)$^{+0.41}_{-0.37}$
($2\sigma$), which denotes that the coincidence problem becomes more
severer with inclusion of the SL test. Moreover, the phantom-like
dark energy ($w_X < -1$) is slightly favored over the $\Lambda$CDM
model. The interaction is thus reconstructed in Figure
\ref{interaction} with a current value
$\delta_0=-0.02^{+0.17}_{-0.19}$ ($1\sigma$). We find that the
best-fit interaction is negative but with relatively large errors
for low redshift, which hints the energy transfer from matter to
dark energy. We also note that the interaction $\delta(z)$ decreases
with the increasing of redshift $z$ and approaches to zero at
infinity.

\section{Conclusion and discussion}
\label{conclusion}

Till now, the constant $\delta$
\cite{wang2005can,amendola2007consequences} and varying coupling
$\delta (z)$ dark energy models \cite{dalal2001testing} have been
revisited using the secular Sandage-Loeb (SL) test. The SL test is
in the inaccessible redshift zone for recent observations, such as
the SNIa, OHD and BAO at $z<2$ and the CMB at $z \simeq 1090$. We
have extended the analysis to the epoch at $2<z<5$ using the secular
redshift drift of the QSO spectra.

For the constant coupling model, the current observation
combinations give a weak interaction term, which is consistent with
previous results \cite{guo2007probing}. By including the simulated
SL test data, we find that they can constrain the corresponding
parameters more stringent, such as the interaction $\delta=-0.01 \pm
0.01 (1\sigma) \pm 0.02
 (2\sigma)$, which has been improved to be only a half of original scale on the errors.
Obviously, the interaction is negative at the $1\sigma$ level. The
joint constraints of SNIa and OHD give a weak constraint on the
interaction. As stated by Guo et al \cite{guo2007probing}, this is
because the CMB data does not allow a large deviation from the
standard matter-dominated epoch, otherwise it can modify the CMB
angular-diameter distance. We extend the analysis to the redshift
interval $2<z<5$ and compare it with combination of SNIa and OHD in
Figure \ref{constant contour2}. We find that the SL test can
constrain the parameters much more stringent. The best-fit of
interaction term is directly reduced from 0.39 to 0.10. So, the
higher-redshift observation including the SL test is necessary to
reveal how the interaction gradually changes with the cosmological
evolution.

For the varying coupling model, a relation $\rho_X \propto \rho_m
a^{\xi}$ is imposed on the density evolution of cosmic components.
Combining the SL test with the current observational data, we find
that they can present more narrowed constraint, which behaves
similar as the constant coupling model. We also reconstruct the
interaction $\delta (z)$ in Figure \ref{interaction}. It is found
that best-fit $\delta (z)$ is negative at low redshift and generally
vanishes for high redshift. Moreover, errors of the reconstructed
$\delta (z)$ are remarkable at low redshift. In this scenario, the
terms $\xi+3w_X=0$ and $\xi+3w_X \neq 0$ respectively denotes the
standard cosmology without interaction and non-standard cosmology.
The $\Lambda$CDM model can be reduced for the case $\xi=3$. In
addition, the coincidence problem will be less severe for the
scaling parameter $0< \xi <3$. From the likelihood test, we find
that the $\Lambda$CDM model still remains a good fit to the recent
observational data and the SL test. That is, the coincidence problem
still exists and is quite severe, which is consistent with previous
results. However, the phantom-like dark energy with $w_X<-1$ is
slightly favored over the $\Lambda$CDM model.

Investigation using the SL test on the constant coupling model shows
that the interaction until redshift $z \sim 5$ still cannot be
neglected. Therefore, it is also reasonable for us to deduce that
the observations at higher redshift, such as the gamma-ray burst may
be useful to detect the interacting model, because some of them even
can be monitored at the redshift $z \sim 8$. We should also note
that the inclusion of  SL test with small sample can narrow the
contour region obtained from large sample with high significance,
which can be evidenced in Figure \ref{constant contour2}.
Furthermore, the high-$z$ SL test is immune from the
model-dependence, calibration of the standard candle, and the
peculiar motion of the observed objects. Therefore, we could expect
that the future SL test will play an important role to test the
cosmological models.

\begin{acknowledgements}

M.J.Zhang would like to thank Zhengxiang Li and Jing-Zhao Qi for
their valuable and kind help on programming. This work was supported
by the National Natural Science Foundation of China (Grant
Nos.11235003, 11175019, 11178007).

\end{acknowledgements}

\bibliographystyle{spphys}
\bibliography{inter}

\end{document}